\documentclass[conference]{IEEEtran}
\IEEEoverridecommandlockouts

\usepackage[utf8x]{inputenc}
\usepackage{amsmath,amssymb,dsfont,stfloats,color,url}
\PassOptionsToPackage{draft}{hyperref}
\usepackage[pdftex]{graphicx}
\usepackage{subfigure}
\usepackage{algpseudocode,algorithm,algorithmicx}
\usepackage{float}
\usepackage{afterpage}
\usepackage{balance}
\usepackage{cite}
\usepackage{array}
\usepackage{verbatim}
\usepackage{comment}

\usepackage{mathbbol}

\def\mindex#1{\index{#1}}

\def\sq{\hbox{\rlap{$\sqcap$}$\sqcup$}}
\def\qed{\ifmmode\sq\else{\unskip\nobreak\hfil
\penalty50\hskip1em\null\nobreak\hfil\sq
\parfillskip=0pt\finalhyphendemerits=0\endgraf}\fi\medskip}

\long\def\defbox#1{\framebox[.9\hsize][c]{\parbox{.85\hsize}{%
\parindent=0pt
\baselineskip=12pt plus .1pt      
\parskip=6pt plus 1.5pt minus 1pt 
 #1}}}

\long\def\beginbox#1\endbox{\subsection*{}%
\hbox{\hspace{.05\hsize}\defbox{\medskip#1\bigskip}}%
\subsection*{}}

\def\endbox{}

\newsavebox{\junk}
\savebox{\junk}[1.6mm]{\hbox{$|\!|\!|$}}

\def\argmin{\mathop{\rm arg\, min}}

\def\K{{\sf K}}

\def\bfmath#1{{\mathchoice{\mbox{\boldmath$#1$}}%
{\mbox{\boldmath$#1$}}%
{\mbox{\boldmath$\scriptstyle#1$}}%
{\mbox{\boldmath$\scriptscriptstyle#1$}}}}

\def\bfmY{\bfmath{Y}}

\def\bfmhhaY{\bfmath{\hhaY}} 
\def\bfmhhaY{\hbox to 0pt{$\widehat{\bfmY}$\hss}\widehat{\phantom{\raise 1.25pt\hbox{$\bfmY$}}}}

\def\til={{\widetilde =}}

 \def\FRAC#1#2#3{\genfrac{}{}{}{#1}{#2}{#3}}

\def\ddtp{{\mathchoice{\FRAC{1}{d^{\hbox to 2pt{\rm\tiny +\hss}}}{dt}}%
{\FRAC{1}{d^{\hbox to 2pt{\rm\tiny +\hss}}}{dt}}%
{\FRAC{3}{d^{\hbox to 2pt{\rm\tiny +\hss}}}{dt}}%
{\FRAC{3}{d^{\hbox to 2pt{\rm\tiny +\hss}}}{dt}}}}

\def\average#1,#2,{{1\over #2} \sum_{#1}^{#2}}

\def\eye(#1){{\bf(#1)}\quad}

\def\eq#1/{(\ref{e:#1})}

\newcommand{\beqn}[1]{\notes{#1}%
\begin{eqnarray} \elabel{#1}}

\newcommand{\eeqn}{\end{eqnarray} }

\newcommand{\beq}[1]{\notes{#1}%
\begin{equation}\elabel{#1}}

\newcommand{\eeq}{\end{equation}}

\def\bdes{\begin{description}}
\def\edes{\end{description}}

\newcounter{rmnum}

\newcounter{anum}

{\end{list}}

\def\ass(#1:#2){(#1\ref{#1:#2})}

\def\ritem#1{
\item[{\sf \ass(\current_model:#1)}]
}

\newenvironment{recall-ass}[1]{%
\begin{description}
\def\current_model{#1}}{
\end{description}
}

\newfont{\bb}{msbm10 scaled 1100}

\newcommand{\dv}{{\bf d}}

\newcommand{\Rm}{{\bf R}}

\newcommand{\Zm}{{\bf Z}}

\newcommand{\Ac}{{\cal A}}
\newcommand{\Bc}{{\cal B}}

\newcommand{\Gc}{{\cal G}}
\newcommand{\Hc}{{\cal H}}

\newcommand{\Sc}{{\cal S}}
\newcommand{\Tc}{{\cal T}}
\newcommand{\Uc}{{\cal U}}

\newcommand{\Vc}{{\cal V}}

\usepackage{hyperref}
\hypersetup{
    bookmarks=true,         
    unicode=false,          
    pdftoolbar=true,        
    pdfmenubar=true,        
    pdffitwindow=false,     
    pdfstartview={FitH},    
    pdfnewwindow=true,      
    colorlinks=true,       
    linkcolor=red,          
    citecolor=green,        
    filecolor=blue,      
    urlcolor=blue           
}

\DeclareGraphicsExtensions{.eps,.pdf,.png,.jpg,.gif,.jpeg,.pstex}

\begin{document}

\title{Routing-Based Delivery in Combination-Type Networks with Random Topology}

\author{\IEEEauthorblockN{Mozhgan~Bayat, Kai~Wan and~Giuseppe~Caire}
\IEEEauthorblockA{ Communications and Information Theory Group, Technische Universit\"{a}t Berlin, 10623 Berlin, Germany}
	\IEEEauthorblockA{E-mails: \{bayat, kai.wan, caire\}@tu-berlin.de}
}
\maketitle

\begin{abstract}
The coded caching scheme proposed by Maddah-Ali and Niesen (MAN) transmits coded multicast messages to users  equipped with caches and it is known to be optimal within a constant factor. 
This work extends this caching scheme to  two-hop relay networks with one main server with access to a library of $N$ files, and $H$ relays communicating with $K$ users with cache, each of which is connected to a random subset of relays. This topology can be considered as a generalized version of a celebrated family of networks, referred to as
{\it combination networks}.
Our approach is simply based on routing MAN packets through the network. The optimization of the routing can be formulated as a Linear Program (LP).
 In addition, to reduce the computation complexity, a dynamic algorithm is proposed to approach the LP solution. Numerical simulations show that the proposed scheme outperforms the existing caching schemes for this class of networks.

\end{abstract}


\section{Introduction}
\label{sec:Introduction}
Due to the growing consumption of on-demand video and audio streaming services with prominent platforms like Youtube, Netflix and Spotify garnering billions of users on a daily basis, a clever usage of the low cost storage to cache data plays a key role in network design. One of the simplest methods is uncoded caching to duplicate popular files at the edge nodes in the networks. The authors in \cite{Golrezaei2012femto} introduced femto caching in heterogeneous wireless networks, where helpers posses high storage and are placed in a fixed position within the cell. The femto base stations cache popular files requested by mobile users. 
Recently, Fog Radio Access Network (F-RAN) has been proposed as a smart network based on  cloud-RAN (C-RAN). In F-RAN the  remote radio heads (RRHs) may posses a local cache as well as  baseband processing units. In the prefetching phase RRHs store popular files in their cache memory. In \cite{shamai2016opt}, the authors study the delivery phase in F-RAN and design the cloud and edge processing jointly. Different transfer fronthaul strategies were proposed in \cite{shamai2016opt} to maximize the delivery rate with limited fronthaul capacities and power constraints.

In parallel, coded caching strategy was orginally proposed by Maddah Ali and Niesen (MAN) in \cite{maddah2014fundamental} for   bottleneck networks with a shared error-free link. In the MAN setting, the server has a library of $N$ files of equal size and is connected to $K$ users through an error-free link. Each of the users is equipped with  a cache memory.
The MAN caching scheme consists of  {\it prefetching} and {\it delivery} phases.
The key idea is to treat the cache content as receiver side information and design the caches during the prefetching phase such that, during the delivery phase, the server can send coded multicast messages from which   users can  retrieve their desired packets with the aid of their own cache content.
The MAN caching scheme which can lead to an additional coded caching gain compared to the conventional uncoded caching scheme, was proved in \cite{ontheoptimality} to be optimal  under the constraint of uncoded prefetching phase when $N \geq K$.
By removing the redundant MAN multicast messages when $N <K$, the authors in \cite{exactrateuncoded} proposed an optimal caching scheme under the constraint of uncoded prefetching phase  
  for any $N$ and $K$.

Recently, the  coded caching approach was applied to different scenarios with specific topologies. \cite{ji2015caching, zewail2017coded,wan2017caching} have investigated a class of two layered symmetric networks, referred to as {\it combination networks }(CN). The topology of these networks in determined by two parameters $H$ and $L$. 
There are a single server that connects to $H$ relays,  and ${H \choose L}$ users each of which is connected to a different $L$-subset of relays.
All links are error-free orthogonal.
 In \cite{ji2015caching} the authors applied the centralized MAN algorithm to create each coded multicast message, which is then encoded with a $(H,L)$ MDS code so that any user that receives $L$ out of the $H$ coded blocks will be able to decode the coded message. 
The server transmit one different MDS coded symbol for each  MAN multicast message to each relay, which then forwards it to the connected users.
In \cite{zewail2017coded}, the authors improved the coded caching scheme in  \cite{ji2015caching} by using the network topology information.
Notice that while the MDS coded scheme of \cite{ji2015caching} applies to any network topology as long as each user can receive from at least $L$ relays, the scheme of \cite{zewail2017coded} critically hinges on the combination network topology and therefore does not generalize to networks with random topology, which is the focus of this work.
In  \cite{wan2017caching} an improvement of \cite{zewail2017coded} obtained by removing redundancy is presented, and in \cite{Llorca2013opt} optimal use of transport and caching resources is achieved by using network coding
on the set of elemental information objects being cached and transported over the network. 

Cache-aided multiple relay networks with random topology was considered  in  \cite{topology2017deniz}, where each user is randomly connected to $L$ relays. By observing that for each relay, there are some MAN multicast messages which are not desired by any of its connected users, the authors  in  \cite{topology2017deniz} presents an improved delivery strategy with respect to \cite{ji2015caching} by using the same MDS coding idea but only transmitting MDS coded symbol to the relays which is connected to at least one user desiring this symbol.

\subsection{Our Contribution}
  This paper considers a    generalized version of the cache-aided two-hop relay networks considered in \cite{topology2017deniz}, where  a server is connected to multiple relays through  unit capacity links and users randomly connect to an identical number of relays. In our work, we consider different capacity links and that the number of connected relays to each user is not necessarily identical. 
 We use the centralized MAN cache placement and in the delivery phase, we first generate coded multicast messages and find the optimal routing for every coded multicast message through {\it Linear Programming} (LP). We also propose a dynamic programming  which approaches the solution of the LP problem  in a recursive manner with much less computational complexity than solving the original LP.

 The distinguishing features of this paper are  two-fold:
\begin{itemize}
	\item We minimize the delivery latency and max-link load of the coded caching scheme through LP. We remove the transmission of non necessary messages and the parity blocks of the MDS coding. 
	\item Our  network topology is a general random relay network with links of different capacity and non-identical number of connected relay to each user. 
	 Even though the proposed LP-based delivery schemes require information about the network's topology, the prefetching phase is independent of the network topology.
\end{itemize}

\section{System Model and Related Results}
\label{sec:SystemModel}

\subsection{System Model}
\label{sub:system}
\paragraph*{\textbf{Notation Convention}}
Calligraphic symbols denote sets, and 
bold symbols denote vectors.
We use
$|\cdot|$ to represent the cardinality of a set;
$[a:b]:=\left\{ a,a+1,\ldots,b\right\}$ and $[n] := [1:n]$.
$\mathcal{A\setminus B}:=\left\{ x\in\Ac : x\notin\Bc\right\}$; 


Our scenario entails a relay network with a random topology and limited capacity error-free links. The server has access to a library consisting of $N$ files $\mathcal{F} = \{W_1, W_2, \dots, W_N\}$ each of which contains $F$ bits. The server is connected to $H$ relays, each of which in turn is serving a random subset of  $K$ users. All links are error-free and orthogonal.  We focus on parallel transmission in which all links works in parallel.
Notice that this is generalized setting of the case where every user is connected to exactly $L$ relays as described in \cite{bayat2017cell, topology2017deniz} and \cite{ji2015caching}. Additionally, each user is equipped with a cache memory capable of storing up to $MF$ bits, while relays do not posses any cache memory.  The subset of users connected to relay $h \in[H]$   and the subset  of relays connected to user $k \in [K]$  are denoted by $\mathcal{U}_h$  and $\mathcal{H}_k$, respectively. Similarly, for each subset of users $\Vc\subseteq [K]$,
we define $\mathcal{H}_{\Vc}= \bigcup_{k\in \Vc}\mathcal{H}_k$ as the union of relays connected to users in $\Vc$. In other words, $\mathcal{H}_{\Vc}$ consists of  relays that connect to at least one user in $\Vc$. 
In this paper, we  consider the system with  limited capacity. The relay nodes are connected to server  via  a fronthaul link has capacity $C_F$ bits per channel use, as well as they are connected to users through links with a given capacity  $C_E$ bits per channel use.

In the prefetching phase, user $k\in[\K]$ stores information about the $N$ files in its cache of size $MF$ bits, where $M \in[0,N]$.  This phase is done without knowledge of users' demands. 
We denote the content in the cache of user $k\in[K]$ by $Z_{k}$ and let $\Zm:=(Z_{1},\ldots,Z_{K})$.

During the delivery phase, user $k\in[K]$ demands file $d_{k}\in[N]$;
the demand vector $\dv:=(d_{1},\ldots,d_{K})$ is revealed to all nodes. 
Given $(\dv,\Zm)$, the server sends a message $X_{s\to h}$ 
of $ R_{h}F$ bits to relay $h\in [H]$. 
Then, relay $h\in [H]$ transmits a message $X_{h\to k}$ 
of $R_{h\to k}F$ bits to user $k \in \Uc_h$. 
User $k\in[K]$ must recover its desired file $F_{d_{k}}$ from $Z_{k}$ and $(X_{h\to k} : h\in \Hc_k)$ with high probability for some $F$. 
The objective is to determine the minimum worst-case transmission time, 
\begin{align}
T^{\star}
:=
\min_{\substack{\Zm}}
\max_{ \dv\in[N]^{K}} \left\{ \max_{h\in [H]} \frac{ R_h F }{C_F}, \max_{h\in [H], k\in \Uc_h} \frac{R_{h\to k} F}{ C_E}  \right\}.
\end{align}

 \begin{figure}[t]
 	\centering
 	\includegraphics[width=0.45\textwidth]{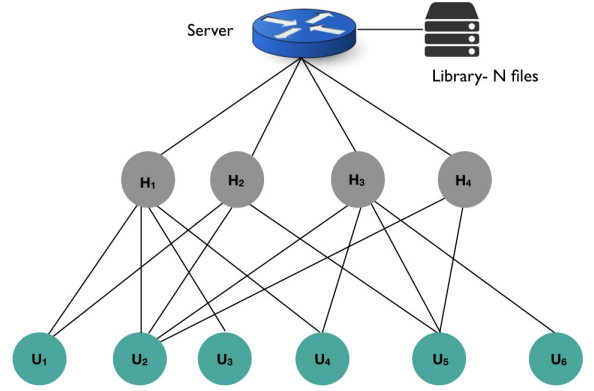}
 	\caption{A random topology with $4$ relays and $6$ users }
 	\label{fig:scenario}
 \end{figure}

\subsection{MAN Caching Scheme}
\label{sub:MAN}

In the following, we introduce the MAN caching scheme for bottleneck caching systems.
 we assume that the library replication parameter $t = KM/N$ (how many times the library can be contained in the collective cache memory) is an integer in $[0:K]$ (for non-integer numbers, the memory sharing is used). Each file $W_i$ is divided into $\binom{K}{t}$ non-overlapping and equal-length subfiles $W_{i}=\{W_{i,\Tc}: \Tc\subseteq [K], |\Tc|=t \}$
with user $k$ caching the segments $W_{i, \Tc}$ for which  $k \in \Tc$. 
In the delivery phase,
The server broadcasts one MAN coded multicast message for each group of users $\Sc$  where $\Sc \subseteq [K]$ and $|\Sc|=t+1$ as
\begin{equation}
\label{eq:coded_multicast}
	V_{\Sc} = \bigoplus_{k \in \Sc} W_{d_k, \Sc\setminus\{k\}}.
\end{equation} 
Each user $k\in \Sc$ requires $W_{d_k, \Sc\setminus\{k\}}$ and knows all other subfiles in $V_{\Sc}$ such that it can recover $W_{d_k, \Sc\setminus\{k\}}$. By considering all the group of users with cardinality $t+1$, each user can recover its desired file.

\section{main result}
\label{sec:MainResult}
\subsection{General Optimization Formulation}
In this section, we introduce our proposed optimization problem for the routing of the coded multicast messages in \eqref{eq:coded_multicast}. 
In order to deliver each MAN  multicast message $V_{\Sc}$, we encode $V_{\Sc}$ into $H$ linearly independent random linear combination messages $X^h_{\Sc}$, $h \in [H]$, with $X^h_{\Sc}$ denoting the coded message corresponding to the relay $h$. In the following step the server transmits $X_{\Sc}^h, h\in[H]$, to the related relays. After receiving $X^h_{\Sc}$, relay $h$ transmits this message to the users in $\Sc \cap \Uc_h$.    
We define the normalized length of the message $X^h_{\Sc}$ as  $y_{\Sc}^h := \frac{|X^h_{\Sc}|}{|V_{\Sc}|}$ where $ 0 \leq y_{\Sc}^h \leq 1$.  We let   $y_{\Sc}^h=0$  when $h \in [H]\setminus \Hc_{\Sc}$; i.e., the coded multicast message $V_{\Sc}$ is delivered only through the relays which are connected to at least one user in $\Sc$.
When the file size $F \rightarrow \infty$, the messages $X_{\Sc}^h, h \in \Hc_k$ are linearly independent with high probability such that 
user $k \in \Sc$ can recover the message $V_{\Sc}$ through these coded messages if
\begin{equation}
\label{eq:cons}
\sum_{ h \in \mathcal{H}_k}  y_S^h \geq 1,  \quad \quad  \forall k \in S.
\end{equation}

The normalized transmission load over the link between the server and each relay $h\in [H]$ is the sum of the loads of all multicast messages and is given by
\begin{equation}
R_h =  \sum_{\Sc \subseteq [K]:|\Sc|=t+1} y_S^h.
\end{equation}
We also define the vector $\Rm_{\mathcal{H} }=\{R_1, R_2, \dots, R_H\}$.
 Similarly, the normalized transmission load between each relay $h\in [H]$ and each user $k\in \Uc_h$ is equal to sum of the loads of all messages as follows
\begin{equation}
R_{h\to k} =  \sum_{\Sc \subseteq [K]:|\Sc|=t+1, k \in {\Sc}} y_{\Sc}^h.  
\end{equation} 

We consider the relay nodes are connected to server  via  a fronthaul link with capacity $C_F$ bits per channel use, as well as they are connected to users though links with a given capacity  $C_E$. We focus on parallel  transmission in which fronthaul links and relay-user links works in parallel. The  delivery time would be maximum of  fronthaul and relays-users delivery time. For given messages $X_{\Sc}^h$ in each block, the required time for   fronthaul transmission can be computed as $\frac{|X_{\Sc}^h|}{C_F}$ , so that the worst-case delivery time in fronthual can be writen as follows 
\begin{align}
T_F  = \frac{\max_{h\in [H]} R_h F }{C_F}  =\max_{h\in[H]}  \sum_{\Sc \subseteq [K]:|\Sc|=t+1 } \frac{y_{\Sc}^h F }{C_F}.
\end{align}
For transmission toward the users the worst-case delivery time from relays to users can be computed as follows
\begin{align}
T_E  &= \max_{h\in [H], k\in \Uc_h} \frac{R_{h\to k} F}{ C_E}  \nonumber\\&= \max_{h\in[H]}\max_{k\in\mathcal{U}_h}  \sum_{\Sc \subseteq [K]:|\Sc|=t+1,  k \in  \Sc} \frac{ y_{\Sc}^h F }{C_E}.
\end{align}
We are interested to minimize the worst-case delivery time by finding optimal random linear combination messages length. 
So that our linear optimization problem can be formalized as follows

 \begin{equation}
 \label{eq:opt_time}
 \begin{aligned}
 &~ \underset{y_{\Sc}^h}{\text{minimize}}
 & &  \max \{T_E ,T_F\}\\
 & ~\text{subject to}
 & & \forall {\Sc},\text{ such that } {\Sc} \subseteq [K], |{\Sc}|= t+1 : \\
 &~ &&  0 \leq y_{\Sc}^h \leq 1,  ~ \text{if}~ h \in \mathcal{H}_{\Sc},  \\
 &~ && y_{\Sc}^h=0,   ~ \text{if}~ h \notin \mathcal{H}_{\Sc}, \\ 
 &~ &&  \sum_{  h \in \mathcal{H}_k}^{} y_{\Sc}^h \geq 1 , ~\forall k\in {\Sc}. \\
 \end{aligned}
 \end{equation}
 \subsection{Optimization Formulation for Identical Capacity Links}
It is straightforward to show that $R_{h\to k} \leq R_h,~ \forall k \in \Uc_h$.
Therefore, the optimization in the case of identical capacity links (for the simplicity, we consider $C_F=1, C_E =1$) reduces to allocating the lengths of the messages $X^h_{\Sc}$, such that the $\max_{h\in[H]} R_h$ is minimized. 
Since for all S, the MAN messages $V_\Sc$ have all the same length, this optimization can be expressed in terms of the normalized loads  $y_{\Sc}^h$ allocation as the following LP

 \begin{equation}
 \label{eq:opt_main}
 \begin{aligned}
 &~ \underset{y_S^h}{\text{minimize}}
 & &  \max_{h\in[H]} R_h \\
 & ~\text{subject to}
 & & \forall {\Sc},\text{ such that } {\Sc} \subseteq [K], |{\Sc}|= t+1 : \\
&~ &&  0 \leq y_{\Sc}^h \leq 1,  ~ \text{if}~ h \in \Hc_{\Sc},  \\
&~ && y_{\Sc}^h=0,   ~ \text{if}~ h \notin \Hc_{\Sc}, \\ 
&~ &&  \sum_{ h \in \Hc_k}^{} y_{\Sc}^h \geq 1 , ~\forall k\in {\Sc}. \\
 \end{aligned}
 \end{equation}
 
 \subsection{Dynamic Programming with Low Complexity}
Each of the LPs in~\eqref{eq:opt_time} and~\eqref{eq:opt_main} contains $\binom{K}{t+1} \times H$ variables and $\binom{K}{t+1} \times (t+1)+H$ constraints, which grow exponentially with the number of users.  For large value of $K$,  it eventually becomes infeasible to solve the optimization problem. In order to overcome this difficulty, we used dynamic programming. In the following, we explain dynamic programming for the  optimization with identical and unit capacities in \eqref{eq:opt_main}. Similarly, the dynamic programming can be easily apply to the worst-case transmission time optimization problem in \eqref{eq:opt_time}.

Consider the optimization problem in~\eqref{eq:opt_main}.
We randomly divide all the $\binom{K}{t+1}$ MAN multicast messages into $G$ non-overlapping and `equal-length' groups.\footnote{If $G$ does not divide $\binom{K}{t+1}$, each of the first $\binom{K}{t+1}-G \lfloor \binom{K}{t+1}/G \rfloor$ groups contains $\lceil \binom{K}{t+1}/G  \rceil $ MAN multicast messages and each of the remaining groups contains $\lceil \binom{K}{t+1}/G  \rceil -1$ MAN multicast messages.} We also let $g:=\lceil \binom{K}{t+1}/G  \rceil $, representing the maximum number of MAN multicast messages included in one group. In addition, the set of the indices of the MAN multicast messages in the $i$-th group is denoted by $\Gc_i$.
The  optimization problem in \eqref{eq:opt_main} can be broken into smaller optimizations by iterating through the groups and using the results from the previous optimization as initial values. 
This allows us to apply minimization of the delivery load  in each group individually in a sequential manner. This means that in the  $i$-th step we take the previous optimal load allocations as a fixed initial load into the next $i+1$-th step's   optimization problem. 
We define the load on relay $h$ in group $\mathcal{G}_i$  as 
\begin{equation}
R_{h,i} =\sum_{\Sc \in \mathcal{G}_i} y_{\Sc}^h.
\end{equation}
 
In the first step, we optimize the load allocation in $\mathcal{G}_1$  with initial normalized load  $\Rm_1^0 = 0$. The initial normalized load for step $i>1$ is obtained by summing all optimal normalized load allocation from previous steps as following
\begin{equation}
R_{h,i}^0 = \sum_{j=1}^{i-1} R_{h,j}^*,
\end{equation}
where $R_{h,j}^*$ is optimal normalized load allocation in step $j$ for relay $h$ and also we denote the vector $\Rm_{\Hc , i}^0 = \{ R_{1,i}^0, R_{2,i}^0, \cdots, R_{H,i}^0\}$ as the initial normalized load allocation burden on relays in the $i$-th step. 
Accordingly, the total normalized load on relay $h$ until step $1 \leq i \leq G$  being equal to $R_{h,i}+R_{h,i}^0$.
Having explained the dynamic programming aspect of Algorithm \ref{alg:dyn}, we will now address the max-link load minimization problem for the $i$-th group with an initial normalized load allocation  $\Rm_{\Hc,i}^0$. Similary, the dynamic programming can be easily apply to the delivery time optimization problem in \eqref{eq:opt_time}.

 \begin{equation}
 \label{eq:greedy_opt}
 \begin{aligned}
 &~ \underset{y_{\Sc}^h}{\text{minimize}}
 & &  \max_{h\in[H]} R_{h,i} +R_{h,i}^0\\
 & ~\text{subject to}
 & & \forall {\Sc},\text{ such that } {\Sc} \in \mathcal{G}_i: \\
 &~ &&  0 \leq y_{\Sc}^h \leq 1,  \\
 &~ && y_{\Sc}^h=0,   ~ \text{if}~ h \notin \mathcal{H}_{\Sc}, \\ 
 &~ &&  \sum_{ h \in \mathcal{H}_k}^{} y_{\Sc}^h \geq 1 , ~\forall k\in {\Sc}. \\
 \end{aligned}
 \end{equation} 
 
Consider that $R_{h,i}^* $ is the optimal value for optimization problem in step $i$. We denote $\Rm_{\mathcal{H},i}^* =\{R_{h,i}^*| h\in[H]\}$ as the total optimal normalized load vector for step $i$.

\begin{algorithm}
	\caption{Optimization with dynamic programming }
	\label{alg:dyn}
	\begin{algorithmic}[1]
		\Require{ geometry of network ,$g$, $G$, $K$, $t$, and $H$  }
		\State Calculate $g:=\lceil \binom{K}{t+1}/G  \rceil $
		\State Divide  the set of all coded multicast messages randomly to $G$ group with g members
		\State Set the initial value $\Rm_{\Hc,i}^0 = \bold{0}$
		\For{$i= 1:G$}
		\State  Solve the optimization problem in \eqref{eq:greedy_opt}
		\State $ \Rm_{\mathcal{H},i}^*= \argmin  \max_{h \in [H]}  R_{h,i} +R_{h,i}^0$ 
		\State Calculate the initial load $R_{H,i+1}^0$ for next optimization 
		\EndFor
	\end{algorithmic}
\end{algorithm}

\section{results and discussions}
\label{sec:NumericalResults}

In this section, we focus on the case that each user is connected to  $L$ relays and compare the worst-case loads or the delivery times achieved by the proposed scheme and the existing schemes. As previously discussed, our scenario is not constrained to have the same number of connections on the user's side. 
 In order to compare  our scheme with previous works based on MDS coding \cite{topology2017deniz,ji2015caching}, we assume that the users have the same number of connections. In our scheme, the transmitted messages to relays have different length but in the works  \cite{topology2017deniz,ji2015caching} the transmitted messages have same length.

\begin{figure}[t]
  	\centering
  	\includegraphics[width=0.45\textwidth]{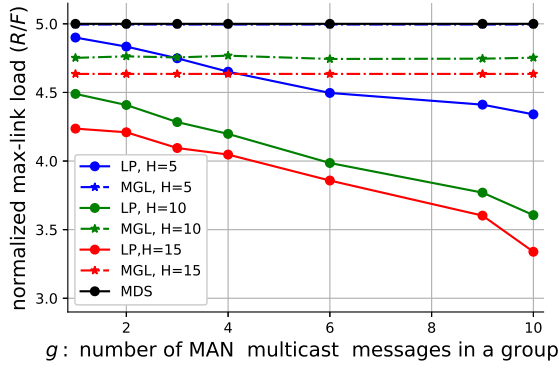}
  	\caption{ The normalized max-link load versus $g$ for a scenario with $K=5$ and $ H=5,10,15$ and $t=2$   }
  	\label{fig:w5}
  \end{figure}

\begin{figure}[t]
  	\centering
  	\includegraphics[width=0.45\textwidth]{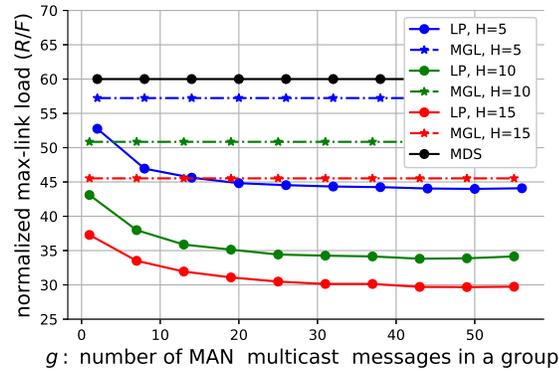}
  	\caption{ The normalized max-link load versus $g$  for a scenario with $K=10$ and $ H=5,10,15$ and $t=2$ with interior point method  }
  	\label{fig:w10}
  \end{figure}

\begin{figure}[t]
  	\centering
  	\includegraphics[width=0.45\textwidth]{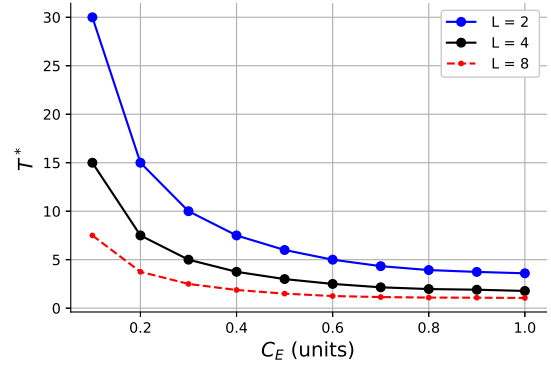}
  	\caption{ The delivery time versus $C_E$ for a scenario with $K=5$ and $ H=10$ and $t=2$ and $C_F = 1 $ unit  }
  	\label{fig:wt}
  \end{figure}

 In all of the figures, the curves labeled ``MDS'' refer to coded caching scheme  for combination networks as described in \cite{ji2015caching}. In ``MDS'' work, first the coded multicast messages $V_{\Sc}$ are created by centralized MAN caching scheme. In the next step, each message is  divided into $L$  subfiles which are then encoded by $(H,L)$ MDS code. Lastly, each of these $H$ MDS coded messages is transmitted to the related relays, which in broadcast the received coded messages to its users. 
The scenario proposed in \cite{topology2017deniz} has been modified to our case by considering that the relays are not equipped with cache memory and curves for this scenario are labeled by ``MGL ''. The ``MGL'' scheme removed some redundancy from \cite{ji2015caching} by avoiding the transmission of coded multicast messages to relays connected to none of users   interested in that specific message.
The curves labeled by ``LP '' refer to the result of our work obtained through the optimization in \eqref{eq:greedy_opt}. 
 The numerical results are obtained by using Monte Carlo simulations that rely on repeated random topologies in which  each user  select L relays at random over all possible relays
 for 500 times. 
 Fig.  \ref{fig:w5} and \ref{fig:w10} show the worst-case load versus $g$ (maximum number of  the MAN multicast messages in one group). Notice that the schemes ``MGL'' and `` MDS'' are independent from $g$. 
As   seen from these numerical results, compared to the previous schemes our scenario dramatically reduces the worst-case load. For an increasing $g$ each optimization in iterative method will include more messages and as a consequence the load association of the relays performs better.
In Fig. \ref{fig:wt} we compare the delivery time of the proposed scheme for different capacities $C_E$ and number of connections on the user side $L$ with a unit fronthaul capacity $C_F = 1$. For smaller capacities, the relay-user links are the bottleneck of our scenario. After increasing $C_E$ up to a certain value, the server-relay links become the bottleneck of our scenario.


\begin{figure}[t]
  	\centering
  	\includegraphics[width=0.45\textwidth]{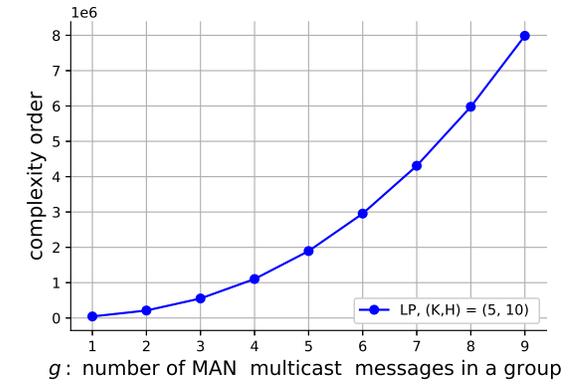}
  	\caption{ Complexity order of whole scenario optimization with interior point method  }
  	\label{fig:complexity}
  \end{figure}

  We used the simplex and interior point methods to solve the optimization problems.  The complexity of simplex on average is  $O(n^3)$  and $O(n^2 2^n)$ in the worst-case, while the complexity of the interior point method is $O(n^{3.5})$, where $n$ is the number of variables \cite{klee1970good, papadimitriou1998combinatorial}. The complexity order of the interior point method for a scenario is shown in Fig. \ref{fig:complexity}. 
   
\section{Conclusion}
In this paper we investigated relay networks where the end users are equipped with caches. We extended the MAN caching scheme for the bottleneck networks to any cache-aided relay networks. For different scenarios with and without unit/identical link capacities, we design a delivery scheme by solving linear optimization problems.
It was concluded in  the numerical results that the proposed schemes outperform the state-of-the-arts schemes in both scenarios.  In addition,
we also proposed  a dynamic algorithm to approach the solution of each optimization problem with a lower computation complexity.


{\small
\bibliographystyle{IEEEtran}
\bibliography{references}
}

\end{document}